\documentclass{appolb}
\usepackage{epsfig}

\begin{document}

\title{ Counting Stationary Points of Random Landscapes as a Random Matrix Problem}
\author{Yan V. Fyodorov
\address{School of Mathematical Sciences,
University of Nottingham , Nottingham NG72RD, United Kingdom}
\address{\small on leave from: Petersburg Nuclear Physics Institute
RAS, 188350 Gatchina, Leningrad Reg., Russia} }
\maketitle

\begin{abstract}
Finding the mean of the total number $N_{tot}$ of stationary
points for $N-$dimensional random Gaussian landscapes can be
reduced to averaging the absolute value of characteristic
polynomial of the corresponding Hessian. First such a reduction is
illustrated for a class of models describing energy landscapes of
elastic manifolds in random environment, and a general method of
attacking the problem analytically is suggested. Then the exact
solution to the problem\cite{my} for a class of landscapes
corresponding to the simplest, yet nontrivial "toy model" with $N$
degrees of freedom is described. For $N\gg 1$ our asymptotic
analysis reveals a phase transition at some critical value $\mu_c$
of a control parameter $\mu$ from a phase with finite landscape
complexity: $N_{tot}\sim e^{N\Sigma},\,\, \Sigma(\mu<\mu_c)>0$ to
the phase with vanishing complexity: $\Sigma(\mu>\mu_c)=0$. This
is interpreted as a transition to a  glass-like state of the
matter.

\end{abstract}

Characterising geometry of a complicated landscape is an important
problem motivated by numerous applications in physics, image
processing and other fields of applied mathematics \cite{Adler}.
The simplest, yet non-trivial task is to find the mean number of
all stationary points (minima, maxima and saddles), which is a
relevant question in statistical physics of disordered (glassy)
systems \cite{sea,halp,toy,spinglass,spinglass1}, and more
recently in string theory\cite{string}.

Assuming that the landscape is described by a sufficiently smooth
random function ${\cal H}$ of $N$ real variables ${\bf
x}=(x_1,...,x_N)$ the problem amounts to finding all solutions of
the simultaneous stationarity conditions  $\partial_k {\cal H}=0$
for all $k=1,...,N$, with $\partial_k$ standing for the partial
derivative $\frac{\partial}{\partial x_k}$. The total number
$N^{(D)}_{tot}$ of its stationary points in any spatial domain $D$
is given by $N_{tot}^{(D)}=\int_D \rho({\bf x}) \, d^N{\bf x}$,
with $\rho({\bf x})$ being the corresponding density of the
stationary points. The ensemble-averaged value of such a density
can be found according to the so-called Kac-Rice formula as
\begin{equation}\label{KR}
\rho_{av}({\bf x})=\left\langle |\det{\left(\partial^2_{k_1,k_2}
{\cal H}\right)}| \prod_{k=1}^N\delta(\partial_k {\cal
H})\right\rangle
\end{equation}
where $\delta(x)$ stands for the Dirac's $\delta-$ function.

Whenever the underlying physical problem necessitates to deal with
an absolute value of the determinant, its presence considered to
be a serious technical challenge, see \cite{Kurchan} and
references therein. In particular, an intensive work and
controversy persists in calculating the so-called  thermodynamic
complexity of the free energy for the standard
Sherrington-Kirkpatrick model of spin glasses\cite{spinglass1} or
its generalisations\cite{spinglass}. Several heuristic schemes
based on various versions of the replica trick were proposed in
the literature recently to deal with the problem, see discussion
and further references in \cite{spinglass1}. Despite some
important insights, the present status of the methods is not yet
completely satisfactory. On the other hand, keeping absolute value
of the determinant is instrumental for dealing with the systems
whose energy function displays many extrema in the configuration
space. A famous example is the notorously difficult case of the
random field Ising model, where disregarding the absolute value
induces an additional symmetry leading to the so-called
"dimensional reduction" prediction \cite{PS}, which proved to be
wrong\cite{Imbrie}.

In this paper we are specifically addressing the problem of
stationary point counting for energy functionals describing an
elastic $d-$dimensional manifold embedded in a random media of
$N+d$ dimensions, see Fig.1. This may serve either as as a model
of domain walls in the random field Ising model \cite{Villain}, or
it is directly related to directed polymers in a random
environment.

\begin{figure}[h!]
\begin{center}
\setlength{\unitlength}{0.75cm}
\begin{picture}(9,7)
\put(0,0){\circle*{0.15},0} \put(2,0){\circle*{0.15},0}
\put(4,0){\circle*{0.15},0} \put(1,1){\circle*{0.15},0}
\put(3,1){\circle*{0.15},0} \put(5,1){\circle*{0.15},0}
\put(2,2){\circle*{0.15},0} \put(4,2){\circle*{0.15},0}
\put(6,2){\circle*{0.15},0} \put(1,4){\vector(0,1){3}}
\put(1,1){\line(0,1){2.475}} \put(0,0){\line(1,1){2}}
\put(2,0){\line(1,1){2}} \put(4,0){\line(1,1){2}}
\put(0,0){\line(1,0){4}} \put(1,1){\line(1,0){4}}
\put(2,2){\line(1,0){4}}        \put(1.25,6.75){$X$}
\qbezier(0,3)(1, 4)(4,3)        \qbezier(4,3)(7, 2)(8,3)
\qbezier(1,5)(2, 6)(5,5)        \qbezier(5,5)(8, 4)(9,5)
\qbezier(0,3)(0, 3.5)(1,5)      \qbezier(8,3)(8, 3.5)(9,5)
\put(2.95,0.55){$i$}    \put(4,2.2){$j$} \put(2.3,2.3){$X_i$}
\put(3,1){\dashbox{0.15}(0.01,3){}} \put(3,4.2){\vector(0,1){0}}
\put(3,4.2){\circle*{0.125},0}\put(3,1){\vector(0,-1){0}}
\end{picture}
\caption{ Sketch of an elastic manifold for dimensions d=2, N=1.}
\end{center}
\end{figure}
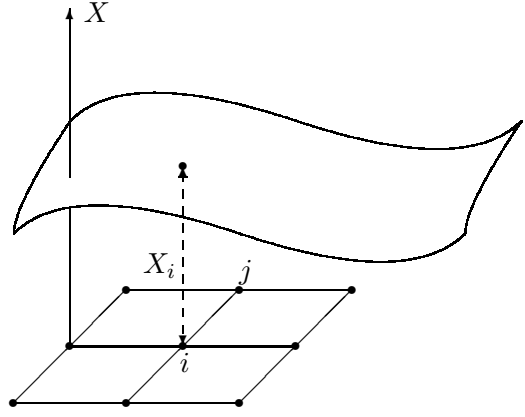

 In a discretized (lattice) version, the simplest form
of the energy functional can be written as
\begin{equation}\label{1}
{\cal H}=\frac{\mu}{2}\sum_{i}{\bf X}^2_i+
\frac{t}{2}\sum_{<i,j>}({\bf X}_i-{\bf X}_j)^2+\sum_{i}
V_i\left[{\bf X}_i\right],\quad \mu\ge 0,t\ge 0.
\end{equation}
Here the index $i$ serves to number points in a d-dimensional
lattice approximating the internal space of the elastic manifold,
and the  $N-$component vectors ${\bf X_i}=\left(X_i^{(1)},\ldots,
X_i^{(N)}\right)$ parametrize transverse deviations of the
manifold from its reference position ${\bf X}_i=0$ at a given site
of the lattice. The first term in (\ref{1}) is a parabolic
confining potential serving to stabilize the reference plane
$X_i=0$, second term stands for the elastic energy interaction
between the nearest neighbours in the lattice, and the last one is
considered to be a random Gaussian function of both $i$ and $X$,
with zero mean and the variance specified by the pair correlation
function chosen in the form:
\begin{equation}\label{2}
\left\langle V_i\left[{\bf X}_i\right] \, V_j\left[{\bf
X}_j\right]\right\rangle=NV^2\delta_{ij}f\left(\frac{1}{2N}({\bf
X}_i-{\bf X}_j)^2\right)
\end{equation}
where $f(x)$ is any smooth function, with suitable decay at
infinity. The stationarity conditions then amount to the set of
simultaneous equations:
\begin{equation}\label{3}
\frac{\partial {\cal H}}{\partial X_i^{(l)}}=\mu
X_i^{(l)}+t\sum_{j}(X_i^{(l)}-X_j^{(l)})+\frac{\partial
V_i}{\partial X_i^{(l)}}=0,\quad l=1,\ldots,N
\end{equation}
so that the mean total number of the stationary points of ${\cal
H}$ amounts according to the Kac-Rice formula to
\begin{equation}\label{4}
\left\langle {\cal N}_{tot}\right\rangle =
\int_{-\infty}^{\infty}\prod_i\prod_{l=1}^N dX_i^{(l)} \left
\langle\delta\left( \frac{\partial {\cal H}}{\partial
X_i^{(l)}}\right)\right\rangle
\end{equation}
$$
\times \left
\langle|\det{\left(\mu\delta_{ij}\delta_{l_1l_2}+t\delta_{l_1l_2}\Delta_{ij}+
\delta_{ij}\frac{\partial^2 V_i}{\partial X_i^{(l_1)}\partial
X_i^{(l_2)}}\right)}|\right\rangle
$$
In the above expression we introduced the matrix $\Delta_{ij}$ for
a discrete Laplacian operator $\Delta$ corresponding to the
underlying lattice and defined via:
\begin{equation}
\Delta_{mn}=\frac{\partial^2}{\partial X_m\partial
X_n}\left[\frac{1}{2}\sum_{<i,j>}(X_i-X_j)^2\right]
\end{equation}
and used that the first and the second derivatives of $V[{\bf X}]$
are (locally) statistically independent due to Gaussian nature of
the disorder, so that the two factors in the integrand can be
averaged independently, which is an immense simplification. In
fact, the averaging of the product of $\delta-$functions in
(\ref{4}) is a rather straightforward task, using independence of
the random potential in different lattice sites and its overall
Gaussian nature. Moreover, the mean value of the modulus of the
determinant is apparently independent of the coordinates $
X^{(l)}_i$ which allows to perform the integration explicitly.
After necessary manipulations, we arrive at an important
intermediate expression
\begin{equation}\label{4}
\left\langle {\cal N}_{tot}\right\rangle =\frac{\left
\langle|\det{\left[\delta_{ij}\left(\mu\delta_{l_1l_2}+\frac{\partial^2
V_i}{\partial X_i^{(l_1)}\partial X_i^{(l_2)}}\right)+
t\delta_{l_1l_2}\Delta_{ij}
\right]}|\right\rangle}{\det^N{\left(\mu\delta_{ij}+t\Delta_{ij}\right)}}
\end{equation}
To proceed further one should be able to calculate the ensemble
average of the modulus of the characteristic determinant of a
lattice Laplacian $\Delta$ perturbed by a kind of random
potential. This problem in full generality is apparently rather
difficult, and we will only discuss here a promising approach
inspired by an analogy with the field of the Anderson localization
of a quantum particle by static disorder. Indeed, in the latter
field the problem of finding the ensemble average of products of
one-particle Green's functions of the lattice operators with
disorder was most successfully tackled in the framework of the
so-called supersymmetry approach due to Efetov\cite{Efetov}, see
also a recent review article by Zirnbauer\cite{Martin}. Following
this analogy, we suggest a technique of evaluating the absolute
value of the determinants for a real symmetric matrix $\mu+H$ via
the following useful identity\cite{my,note}:
\begin{eqnarray}\label{absreg}
|\det{(\mu+H)}|=\lim_{\epsilon\to 0}
\frac{[\det{(\mu+H)}]^2}{\sqrt{\det{\left(\mu+i\epsilon+H\right)}}
\sqrt{\det{\left(\mu+i\epsilon+H\right)}}}
\end{eqnarray}
 The two factors in the denominator of the right
hand side of (\ref{absreg}) can be represented in terms of the
gaussian integrals absolutely convergent as long as $\epsilon>0$.
Further representing the determinantal factors in the numerator in
terms of the Gaussian integral over anticommuting (Grassmann)
variables we thus get a {\it bona fide}
supersymmetric\cite{Efetov} object to be analysed. Actual
implementation of this method is far from being trivial and we
leave it as an interesting problem for future investigations.
Nevertheless, it is important to stress that a possibility to
perform the ensemble average explicitly exists whenever matrix
entries of $H$ are Gaussian-distributed. Similar strategy may be
even employed when $H$ is a stochastic differential operator with
certain Gaussian part, as in the notorously difficult case of the
random field Ising model \cite{PS}. For this reason it is natural
to hope that the suggested method may appear of certain utility
for a broad class of disordered models.

In the rest of the present paper we discuss briefly an explicit
and rigorous solution \cite{my} of the counting problem for a
simple, yet nontrivial limiting case of a "zero-dimensional"
manifold (d=0) whose energy functional is of the form
\begin{equation}\label{fundef}
{\cal H}=\frac{\mu}{2}\sum_{k=1}^N x_k^2+V(x_1,...,x_N)
\end{equation}
with a random Gaussian-distributed potential $V({\bf x})$
characterized by a particular pair correlation function $f$ as in
(\ref{2}). This expression follows from (\ref{1}) after neglecting
the elastic coupling: $t=0$, and as such just serves to describe
behaviour of a single particle in a random potential. Despite
seeming simplicity, the problem of studying this "toy
model"\cite{Villain} dynamics and thermodynamics is still rather
non-trivial, with many features typical of complex, glassy
systems. In particular, it is known to display a very nontrivial
glassy behaviour at low enough temperatures - an unusual
off-equilibrium relaxation dynamics attributed to a complex
structure of the energy landscape. Although particular dynamical
as well as statical properties may differ substantially for
different functions $f(x)$ (e.g. "long-ranged" vs. "short-range"
correlated potentials, see \cite{CKD}), the very fact of glassy
relaxation is common to all of them. In fact, the same model
admits an alternative interpretation as a spin-glass, with $x_i$
being looked at as "soft spins" in a quadratic well interacting
via a random potential $V$\cite{MP}. From this point of view it is
most interesting to concentrate on the limit of large number of
"spins" $N\gg 1$. The experience accumulated from working with
various types of spin-glass models\cite{spinglass} suggests that
for the energy landscape to be complex enough to induce a glassy
behaviour the total number of stationary points $N_{tot}(\mu)$
should grow exponentially with $N$ as
$N_{tot}(\mu)\sim\exp{N\Sigma(\mu)}$. The quantity $\Sigma(\mu)>0$
in such a context is natural to call the {\it landscape
complexity}. On the other hand, it is completely clear that the
number of stationary points should tend to $N_{tot}=1$ for very
large $\mu$ when the random part is negligible in comparison with
the deterministic one. In fact, when $N\to \infty$ we will find
that a kind of sharp transition to the phase with vanishing
complexity occurs at some finite critical value $\mu_{c}$, so that
$\Sigma(\mu)=0$ as long as $\mu>\mu_{c}$, whereas $\Sigma(\mu)>0$
for $\mu<\mu_{c}$ and tends to zero quadratically when
$\mu\to\mu_c$. Such a transition is just the glass transition
observed earlier in a framework of a different approach in
\cite{MP,CKD}.

 For small $N=1,2$
statistics of stationary points of (\ref{fundef}) were
investigated long ago in a classical study of specular light
reflection from a random sea surface\cite{sea} and addressed
several times ever since in various physical contexts, see
\cite{halp,toy}. Let demonstrate that the case of arbitrary $N$
can be very efficiently studied by reducing it to a problem
typical for the well-developed theory of large random
matrices\cite{Mehta}. Indeed, adopting formula (\ref{4}) to the
present case we see that the total number of stationary points in
the whole space is given by:
\begin{equation}\label{tot}
N_{tot}(\mu)=\frac{1}{\mu^N}\left\langle|\det(\mu-H)|\right\rangle
\end{equation}
where $-\hat{H}$ stands for the matrix of second derivatives of
the potential: $H_{k_1k_2}\equiv \partial^2_{k_1,k_2} V$. We see
that the problem basically amounts to evaluating the ensemble
average of the absolute value of the characteristic polynomial
$\det(\mu I_N-H)$ (a.k.a. spectral determinant)  of a particular
random matrix $H$. statistical properties of the potential $V$
result in the following second-order moments of the entries
$H_{ij}\, \,\{(i,j)=1,...,N\}$:
\begin{equation}
\left\langle H_{il}H_{jm}\right\rangle=\frac{J^2}{N}\left[
\delta_{ij}\delta_{lm}+\delta_{im}\delta_{lj}+
\delta_{il}\delta_{jm}\right]
\end{equation}
where we denoted $J^2=f^{''}(0)$.  This allows one to write down
the density of the joint probability distribution (JPD) of the
matrix $H$ explicitly as
\begin{equation}\label{JPD}
{\cal P}(H)dH\propto dH\exp\left\{-\frac{N}{4J^2}
\left[\mbox{Tr}\left(H^2\right)-\frac{1}{N+2} \left(\mbox{Tr}
H\right)^2\right]\right\}
\end{equation}
where $dH=\prod_{1\le i\le j\le N} dH_{ij}$ and the
proportionality constant can be easily found from the
normalisation condition and will be specified later on. It is
evident that such a JPD is invariant with respect to rotations
$H\to O^{-1}HO$ by orthogonal matrices $O\in O(N)$, but it is
nevertheless different from the standard one typical for the
so-called Gaussian Orthogonal Ensemble (GOE)\cite{Mehta}. However,
introducing one extra Gaussian integration it is in fact
straightforward to relate averaging over the JPD (\ref{JPD}) to
that over the standard GOE. In particular,
\begin{eqnarray}\label{starting}
\left\langle|\det{(\mu -H)}|\right\rangle=\int_{-\infty}^{\infty}
\frac{dt}{\sqrt{2\pi}}e^{-N\frac{t^2}{2}}
\left\langle|\det{\left[(\mu+Jt)-H_0\right]}|\right\rangle_{GOE}
\end{eqnarray}
where the averaging over $H_0$ is performed with the GOE-type
measure:
$dH_0\,C_N\exp\left\{-\frac{N}{4J^2}\mbox{Tr}H_0^2\right\}$, with
$C_N=N^{1/2}/[(2\pi J^2/N)^{N(N+1)/4}2^{N/2}]$ being the relevant
normalisation constant.

To evaluate the ensemble averaging in (\ref{starting}) in the most
economic way one
 can exploit explicitly the mentioned rotational $O(N)-$invariance
and at the first step in a standard way\cite{Mehta} reduce the
ensemble averaging to the integration over eigenvalues
$\lambda_1,...,\lambda_N$ of the matrix $H_0$. After a convenient
rescaling $\lambda_i\to J \sqrt{2/N}\lambda_i$ the resulting
expression acquires the following form:
\begin{eqnarray}
&&\left\langle|\det{\left[(\mu+Jt)-H_0\right]}|\right\rangle_{GOE}\\
\nonumber&\propto& \int_{-\infty}^{\infty}d\lambda_1\ldots
\int_{-\infty}^{\infty} d\lambda_N
\prod^N_{i<j}|\lambda_i-\lambda_j| \prod_{i=1}^N
|\sqrt{N/2}(m+t)-\lambda_i| e^{-\frac{1}{2}\lambda_i^2}
\end{eqnarray}
where we denoted $m=\mu/J$. One may notice that the above $N-$fold
integral can be further rewritten as a $N+1$ fold integral:
\begin{eqnarray}
&& e^{\frac{N}{4}(m+t)^2} \int_{-\infty}^{\infty}d\lambda_1\ldots
\int_{-\infty}^{\infty} d\lambda_{N+1} \\\nonumber &\times&
\prod_{i=1}^{N+1} e^{-\frac{1}{2}\lambda_i^2}
\delta\left(\sqrt{N/2}(m+t)-\lambda_{N+1}\right)
\prod^{N+1}_{i<j}|\lambda_i-\lambda_j|
\end{eqnarray}
Such a representation makes it immediately evident that, in fact,
the expectation value of the modulus of the determinant in
question is simply proportional to the mean spectral density
$\nu_{N+1}[m+t]$ (a.k.a one-point correlation function
$R^{(N+1)}_{1}\left[\sqrt{N/2}(m+t)\right]$, see \cite{Mehta}) of
the same GOE matrix $H_0$ but of enhanced size $(N+1)\times
(N+1)$:
\begin{eqnarray}\label{den}
&&\left\langle|\det{\left[(\mu+Jt)I_N-H_0\right]}|\right\rangle_{GOE}
\propto e^{\frac{N}{4}(m+t)^2} \nu_{N+1}[(m+t)],\\ \nonumber &&
\nu_N[\lambda]=\frac{1}{N}\left\langle \mbox{Tr}\,
\delta(\lambda\, I_N-H_0)\right\rangle_{GOE}.
\end{eqnarray}
The last relation provides the complete solution of our original
problem for any value of $N$, since the one-point function
$R^{(N+1)}_{1}\left[x\right]$  is known in a closed
form\cite{Mehta} for any value of $N$ in terms of the Hermite
polynomials $H_k(x)$.

Being interested mainly in extracting the complexity
$$\Sigma(\mu)=\lim_{N\to\infty}N^{-1}\ln{N_{tot}(\mu)},$$ we have to
perform an asymptotic analysis of the mean eigenvalue density, and
substitute the resulting expression into the integral
(\ref{starting}). In the latter we can exploit the saddle-point
method for asymptotic analysis. For $0<m<1$ the relevant saddle
point is $t_s=m$ satisfying $0<\lambda_s=t_s+m<2$, and validating
the use of the semicircular spectral density $\nu[\lambda_s]=
\frac{1}{2\pi}\sqrt{4-\lambda_s^2}$ in the calculation. This
yields
\begin{eqnarray}\label{fin1}
\left\langle|\det{(\mu I_N-H)}|\right\rangle\propto
e^{\frac{N}{2}\,(m^2-1)}\sqrt{1-m^2},\quad 0<m<1
\end{eqnarray}
For $m>1$, however, it turns out that one has to use
 exponentially small
("instanton")  value for the spectral density:
\begin{equation}\label{instanton}
\nu[\lambda]\propto
\exp\left\{-N\left[\frac{1}{4}\lambda\sqrt{\lambda^2-4}
-\ln{\frac{\lambda+\sqrt{\lambda^2-4}}{2}}\right]\right\}, \quad
\lambda>2
\end{equation}
where we only kept factors relevant for calculating the complexity
in the limit of large $N$.
 The corresponding
saddle-point value $t_s$ in the $t-$integral
 is given by the solution of the equation
$m=\frac{1}{2}(\lambda_s+\sqrt{\lambda_s^2-4})$ for the variable
$\lambda_s=t_s+m$. The solution is easily found to be simply
$\lambda_s=m+m^{-1}$ (note that $\lambda_s>2$ ensuring consistency
of the procedure) which yields the resulting value for the modulus
of the determinant to be given by
\begin{eqnarray}\label{fin2}
\left\langle|\det{(\mu I_N-H)}|\right\rangle\propto
e^{N\ln{m}},\quad m>1
\end{eqnarray}
Invoking our basic relation Eq.(\ref{tot}) for $N_{tot}$ we see
that the landscape complexity $\Sigma(\mu)$ of the random
potential function (\ref{fundef}) is given by:
\begin{eqnarray}\label{complexity}
&&\Sigma(\mu)=\frac{1}{2}\,
\left(\frac{\mu^2}{J^2}-1\right)-\ln{\left(\mu/J\right)},\quad
\mu<\mu_{c}=J\\ && \Sigma(\mu)=0,\quad \mu>\mu_{c}=J
\end{eqnarray}
Earlier works referred to the critical value $\mu_{c}=J$ as, on
one hand, signalling the onset of a nontrivial glassy
dynamics\cite{CKD}, and, on the other hand, corresponding to the
point of a breakdown of the replica-symmetric solution\cite{MP}.
Our calculation provides an independent support to the point of
view attributing both phenomena to extensive number of stationary
points in the energy landscape. At the critical value the
complexity vanishes quadratically: $\Sigma(\mu\to \mu_c)\propto
(\mu_c-\mu)^2/\mu_{c}^2$.

The author is grateful to Steve Zelditch for bringing interesting
reference \cite{string} to his attention.
 The work was supported by EPSRC grant EP/C515056/1
"Random Matrices and Polynomials: a tool to understand
complexity".



\end{document}